# $D^0 \to K^\pm \pi^\mp$ and $CP$ Violation at a $\tau$-charm Factory

Zhi-zhong Xing [1]

*Sektion Physik, Theoretische Physik, Universität München,*
*Theresienstrasse 37, D-80333 München, Germany*

**Abstract**

We calculate the time-dependent and time-integrated decay rates of $(D^0 \bar{D}^0) \to (K^\pm \pi^\mp)(l^\pm X^\mp)$ at the $\psi(3.77)$ and $\psi(4.14)$ resonances. The possibilities to distinguish between the effects of $D^0 - \bar{D}^0$ mixing and doubly Cabibbo suppressed decay (DCSD) are illustrated, and the signal of $CP$ violation induced by the interplay of mixing and DCSD is discussed. Ratios of the decay rates of $(D^0 \bar{D}^0) \to (K^\pm \pi^\mp)(K^\pm \pi^\mp)$ to that of $(D^0 \bar{D}^0) \to (K^+ \pi^-)(K^- \pi^+)$ are also recalculated by accommodating $CP$ violation and final-state interactions.

---

[1] Electronic address: Xing@hep.physik.uni-muenchen.de

# 1. Introduction

Recently some attention has been paid to the potential of searching for large $D^0 - \bar{D}^0$ mixing that is out of reach of the standard-model limitation, and to the possibility of probing significant $CP$ violation in the charm sector [1, 2, 3, 4, 5]. The main experimental scenarios include the fixed target facilities, the $\tau$-charm factories and the $B$-meson factories [6, 7, 8, 9]. Among various decay channels of neutral $D$ mesons, $D^0$ vs $\bar{D}^0 \to K^\pm \pi^\mp$ are of particular interest, in both theory and experiments, to study the effects of $D^0 - \bar{D}^0$ mixing and doubly Cabibbo suppressed decay (DCSD).

In the standard model, the transitions $D^0 \to K^- \pi^+$ and $\bar{D}^0 \to K^+ \pi^-$ are Cabibbo favored. In contrast, $D^0 \to K^+ \pi^-$ and $\bar{D}^0 \to K^- \pi^+$ are DCSD's. If there exists detectable $D^0 - \bar{D}^0$ mixing due to new physics, then the processes $D^0 \to \bar{D}^0 \to K^+ \pi^-$ and $\bar{D}^0 \to D^0 \to K^- \pi^+$ may compete with or even dominate over the respective DCSD's. Since any new physics is not likely to affect the direct decays of $c$ quark (via the tree-level $W$-mediated diagrams) in a significant way [5, 10], one can use $D^0$ vs $\bar{D}^0 \to K^\pm \pi^\mp$ to explore the magnitude of $D^0 - \bar{D}^0$ mixing as well as the effect of $CP$ violation induced by the interplay of decay and mixing. Of course, this idea has been extensively considered in the literature (see, e.g., refs. [1, 5]). But most of the previous studies focused on the *incoherent* decays of $D^0$ and $\bar{D}^0$ mesons, a case applicable to the fixed target experiments (perhaps also applicable to the experiments at a $B$-meson factory) [6, 9].

In this work we calculate the time-dependent and time-integrated transition probabilities of *coherent* $(D^0 \bar{D}^0)$ decays to $(K^\pm \pi^\mp)(l^\pm X^\mp)$, where the semileptonic final states serve to tag the flavors of the parent $D$ mesons, for various possible measurements at a $\tau$-charm factory. We illustrate how to distinguish between $D^0 - \bar{D}^0$ mixing and DCSD effects in the time distributions of decay rates, and how to isolate the signal of $CP$ violation arising from the interplay of these two effects. Accommodating $CP$ violation and nonvanishing strong phase shift in $D \to K\pi$, a recalculation of the coherent decays $(D^0 \bar{D}^0) \to (K^\pm \pi^\mp)(K^\pm \pi^\mp)$ is also given to probe $D^0 - \bar{D}^0$ mixing and to separate it from the DCSD effect. Similar analyses can be carried out for a variety of neutral $D$ decays to non-$CP$ eigenstates.

# 2. Master formulas

At a $\tau$-charm factory, the coherent $(D^0 \bar{D}^0)$ events can be produced through [7]

$$
\begin{aligned}
e^+ e^- &\to \psi(3.77) \to (D^0 \bar{D}^0)_{C=-} \,, \\
e^+ e^- &\to \psi(4.14) \to \gamma (D^0 \bar{D}^0)_{C=+} \text{ or } \pi^0 (D^0 \bar{D}^0)_{C=-} \,,
\end{aligned}
$$

where $C$ represents the charge-conjugation parity. Since $D^0$ and $\bar{D}^0$ mix coherently until one of them decays, one may use the semileptonic decay of one meson to tag the flavor of the other meson decaying to a flavor-nonspecific hadronic state. The time-dependent wave function for a $(D^0\bar{D}^0)_C$ pair at rest is written as

$$\frac{1}{\sqrt{2}}\left[|D^0(\mathbf{k},t)\rangle \otimes |\bar{D}^0(-\mathbf{k},t)\rangle \; + \; C|D^0(-\mathbf{k},t)\rangle \otimes |\bar{D}^0(\mathbf{k},t)\rangle\right] \; , \tag{1}$$

where $\mathbf{k}$ is the three-momentum vector of $D^0$ and $\bar{D}^0$ mesons. The proper-time evolution of an initially ($t=0$) pure $D^0$ or $\bar{D}^0$ is given by

$$\begin{aligned}
|D^0(t)\rangle &= f_+(t)|D^0\rangle \; + \; e^{+\mathrm{i}2\phi_m}f_-(t)|\bar{D}^0\rangle \; , \\
|\bar{D}^0(t)\rangle &= f_+(t)|\bar{D}^0\rangle \; + \; e^{-\mathrm{i}2\phi_m}f_-(t)|D^0\rangle \; ,
\end{aligned} \tag{2}$$

where $\phi_m$ is a complex parameter connecting the flavor eigenstates to the mass eigenstates through [2]

$$\begin{aligned}
|D_\mathrm{L}\rangle &= e^{-\mathrm{i}\phi_m}|D^0\rangle \; + \; e^{+\mathrm{i}\phi_m}|\bar{D}^0\rangle \; , \\
|D_\mathrm{H}\rangle &= e^{-\mathrm{i}\phi_m}|D^0\rangle \; - \; e^{+\mathrm{i}\phi_m}|\bar{D}^0\rangle \; ;
\end{aligned} \tag{3a}$$

and the evolution functions $f_\pm(t)$ read

$$f_\pm(t) \; = \; \frac{1}{2}e^{-(\mathrm{i}m+\Gamma/2)t}\left[e^{+(\mathrm{i}\Delta m-\Delta\Gamma/2)t/2} \; \pm \; e^{-(\mathrm{i}\Delta m-\Delta\Gamma/2)t/2}\right] \; . \tag{3b}$$

In the above equation, we have defined $m \equiv (m_\mathrm{L}+m_\mathrm{H})/2$, $\Gamma \equiv (\Gamma_\mathrm{L}+\Gamma_\mathrm{H})/2$, $\Delta m \equiv m_\mathrm{H}-m_\mathrm{L}$ and $\Delta\Gamma \equiv \Gamma_\mathrm{L}-\Gamma_\mathrm{H}$, where $m_\mathrm{L(H)}$ and $\Gamma_\mathrm{L(H)}$ are the mass and width of $D_\mathrm{L(H)}$ respectively.

Now we consider the case that one $D$ meson decays to a semileptonic state $|l^+X^-\rangle$ or $|l^-X^+\rangle$ at (proper) time $t_1$ and the other to $|K^+\pi^-\rangle$ or $|K^-\pi^+\rangle$ at $t_2$. After a lengthy calculation, the joint decay rates for having such events are obtained as

$$\begin{aligned}
\mathrm{R}(l^+,t_1;K^+\pi^-,t_2)_C &\propto |A_l|^2\,|A_{K\pi}|^2\,e^{-\Gamma t_+}\left[(1+|\lambda_{K\pi}|^2)\cosh\left(\tfrac{\Delta\Gamma}{2}t_C\right)\right.\\
&\quad \left.-\,2\mathrm{Re}\lambda_{K\pi}\sinh\left(\tfrac{\Delta\Gamma}{2}t_C\right)+(1-|\lambda_{K\pi}|^2)\cos(\Delta m t_C)-2\mathrm{Im}\lambda_{K\pi}\sin(\Delta m t_C)\right] \; ,
\end{aligned} \tag{4a}$$

$$\begin{aligned}
\mathrm{R}(l^-,t_1;K^-\pi^+,t_2)_C &\propto |A_l|^2\,|A_{K\pi}|^2\,e^{-\Gamma t_+}\left[(1+|\tilde{\lambda}_{K\pi}|^2)\cosh\left(\tfrac{\Delta\Gamma}{2}t_C\right)\right.\\
&\quad \left.-\,2\mathrm{Re}\tilde{\lambda}_{K\pi}\sinh\left(\tfrac{\Delta\Gamma}{2}t_C\right)+(1-|\tilde{\lambda}_{K\pi}|^2)\cos(\Delta m t_C)-2\mathrm{Im}\tilde{\lambda}_{K\pi}\sin(\Delta m t_C)\right] \; ;
\end{aligned} \tag{4b}$$

and

$$\begin{aligned}
\mathrm{R}(l^-,t_1;K^+\pi^-,t_2)_C &\propto |A_l|^2\,|A_{K\pi}|^2\,e^{-4\mathrm{Im}\phi_m}\,e^{-\Gamma t_+}\left[(1+|\lambda_{K\pi}|^2)\cosh\left(\tfrac{\Delta\Gamma}{2}t_C\right)\right.\\
&\quad \left.-\,2\mathrm{Re}\lambda_{K\pi}\sinh\left(\tfrac{\Delta\Gamma}{2}t_C\right)-(1-|\lambda_{K\pi}|^2)\cos(\Delta m t_C)+2\mathrm{Im}\lambda_{K\pi}\sin(\Delta m t_C)\right] \; ,
\end{aligned} \tag{5a}$$

$$\begin{aligned}
\mathrm{R}(l^+,t_1;K^-\pi^+,t_2)_C &\propto |A_l|^2\,|A_{K\pi}|^2\,e^{+4\mathrm{Im}\phi_m}\,e^{-\Gamma t_+}\left[(1+|\tilde{\lambda}_{K\pi}|^2)\cosh\left(\tfrac{\Delta\Gamma}{2}t_C\right)\right.\\
&\quad \left.-\,2\mathrm{Re}\tilde{\lambda}_{K\pi}\sinh\left(\tfrac{\Delta\Gamma}{2}t_C\right)-(1-|\tilde{\lambda}_{K\pi}|^2)\cos(\Delta m t_C)+2\mathrm{Im}\tilde{\lambda}_{K\pi}\sin(\Delta m t_C)\right] \; .
\end{aligned} \tag{5b}$$

---

[2] Here $CPT$ symmetry in the $D^0-\bar{D}^0$ mixing matrix has been assumed.

In deriving these formulas, we have used the $\Delta Q = \Delta C$ rule and $CPT$ invariance as well as the reliable assumption that there is no direct $CP$ violation in the decay amplitudes of $D^0 \to K^\pm \pi^\mp$. The relevant quantities appearing in eqs. (4) and (5) are defined as follows: $t_C \equiv t_2 + C t_1$, $A_l \equiv \langle l^+ X^- | \mathcal{H} | D^0 \rangle$, $A_{K\pi} \equiv \langle K^- \pi^+ | \mathcal{H} | D^0 \rangle$, $\lambda_{K\pi} \equiv \exp(-2i\phi_m)\, \xi_{K\pi}$ and $\tilde{\lambda}_{K\pi} \equiv \exp(+2i\phi_m)\, \tilde{\xi}_{K\pi}$, where

$$\xi_{K\pi} \equiv \frac{\langle K^+ \pi^- | \mathcal{H} | D^0 \rangle}{\langle K^+ \pi^- | \mathcal{H} | \bar{D}^0 \rangle}, \qquad \tilde{\xi}_{K\pi} \equiv \frac{\langle K^- \pi^+ | \mathcal{H} | \bar{D}^0 \rangle}{\langle K^- \pi^+ | \mathcal{H} | D^0 \rangle}. \tag{6}$$

Note that $|\tilde{\xi}_{K\pi}| = |\xi_{K\pi}|$ holds in the absence of direct $CP$ violation. But in general $\tilde{\xi}_{K\pi} = \xi_{K\pi}^*$ is not true due to the existence of final-state interactions [5]. Moreover, one should keep in mind that $|\xi_{K\pi}|$ is doubly Cabibbo suppressed and its magnitude is of the order $10^{-2}$ [1].

It is known that $\mathrm{Im}\phi_m \neq 0$ implies observable $CP$ violation in $D^0 - \bar{D}^0$ mixing. This kind of $CP$-violating signal can manefest itself in the like-sign dilepton events of $(D^0 \bar{D}^0)_C$ pairs:

$$\begin{aligned} N_C^{--} &\equiv \mathrm{R}(l^-, t_1; l^-, t_2)_C \propto e^{-4\mathrm{Im}\phi_m}\, e^{-\Gamma t_+} \left[ \cosh\left(\frac{\Delta\Gamma}{2} t_C\right) - \cos(\Delta m t_C) \right], \\ N_C^{++} &\equiv \mathrm{R}(l^+, t_1; l^+, t_2)_C \propto e^{+4\mathrm{Im}\phi_m}\, e^{-\Gamma t_+} \left[ \cosh\left(\frac{\Delta\Gamma}{2} t_C\right) - \cos(\Delta m t_C) \right]. \end{aligned} \tag{7}$$

If $\mathrm{Im}\phi_m$ is of the level $O(\geq 10^{-3})$, it should be first observed from the $CP$-violating asymmetry

$$\frac{N_C^{++} - N_C^{--}}{N_C^{++} + N_C^{--}} = \frac{e^{+4\mathrm{Im}\phi_m} - e^{-4\mathrm{Im}\phi_m}}{e^{+4\mathrm{Im}\phi_m} + e^{-4\mathrm{Im}\phi_m}} \approx 4\mathrm{Im}\phi_m. \tag{8}$$

Note that this asymmetry is not only independent of the time distributions of $N_C^{\pm\pm}$ but also independent of the charge-conjugation parity $C$, thus it can be measured by using the time-integrated dilepton events at either $\psi(3.77)$ or $\psi(4.14)$ resonance.

In the following discussions about the decay modes $D^0 \to K^\pm \pi^\mp$ and their $CP$-conjugate processes, we shall neglect the contribution from $\mathrm{Im}\phi$. We shall in turn use the assumption $|\Delta\Gamma| << |\Delta m|$ (i.e., neglecting the mixing effect induced by $\Delta\Gamma$), which is very likely to be a good approximation if $|\Delta m|$ is close to its current experimental bound [5]. As a consequence, $|\tilde{\lambda}_{K\pi}| = |\lambda_{K\pi}|$ holds.

## 3. Time dependence of the decay rates

By use of the approximations mentioned above, the formulas in eqs. (4) and (5) can be simplified. Here we assume that a reconstruction of the decay-time differences $t_-$ between neutral $D$ decays to $l^\pm X^\mp$ and $K^\pm \pi^\mp$ is possible in experiments. Usually it is difficult to detect the $t_+$ distribution of joint decay rates in either linacs or storage rings, since the creation point of the $\psi(3.77)$ or $\psi(4.14)$ resonance cannot be well resolved [11]. Hence we prefer to integrate $\mathrm{R}(l^\pm, t_1; K^\pm \pi^\mp, t_2)_C$ over $t_+$ in order to obtain the $t_-$ distributions. For simplicity, we define a dimensionless parameter $T \equiv \Gamma t_-$ and choose $T > 0$ by convention. The relevant results are given as follows:

$$\begin{align}
\text{R}(l^+, K^+\pi^-; T)_- &\propto e^{-T} \left(4 - x^2 T^2 - 4xT\text{Im}\lambda_{K\pi}\right), \\
\text{R}(l^-, K^-\pi^+; T)_- &\propto e^{-T} \left(4 - x^2 T^2 - 4xT\text{Im}\tilde{\lambda}_{K\pi}\right), \\
\text{R}(l^-, K^+\pi^-; T)_- &\propto e^{-T} \left(4|\lambda_{K\pi}|^2 + x^2 T^2 + 4xT\text{Im}\lambda_{K\pi}\right), \\
\text{R}(l^+, K^-\pi^+; T)_- &\propto e^{-T} \left(4|\tilde{\lambda}_{K\pi}|^2 + x^2 T^2 + 4xT\text{Im}\tilde{\lambda}_{K\pi}\right);
\end{align} \tag{9}$$

and

$$\begin{align}
\text{R}(l^+, K^+\pi^-; T)_+ &\propto e^{-T} \left[4 - x^2(2 + 2T + T^2) - 4x(1+T)\text{Im}\lambda_{K\pi}\right], \\
\text{R}(l^-, K^-\pi^+; T)_+ &\propto e^{-T} \left[4 - x^2(2 + 2T + T^2) - 4x(1+T)\text{Im}\tilde{\lambda}_{K\pi}\right], \\
\text{R}(l^-, K^+\pi^-; T)_+ &\propto e^{-T} \left[4|\lambda_{K\pi}|^2 + x^2(2 + 2T + T^2) + 4x(1+T)\text{Im}\lambda_{K\pi}\right], \\
\text{R}(l^+, K^-\pi^+; T)_+ &\propto e^{-T} \left[4|\tilde{\lambda}_{K\pi}|^2 + x^2(2 + 2T + T^2) + 4x(1+T)\text{Im}\tilde{\lambda}_{K\pi}\right],
\end{align} \tag{10}$$

where $x \equiv \Delta m/\Gamma$ is a $D^0 - \bar{D}^0$ mixing parameter. In obtaining eqs. (9) and (10), we have assumed $x$ to be at the detectable level [12] (i.e., $|x| \sim 10^{-2}$) and made approximations up to the accuracy of $O(x^2)$, $O(|\lambda_{K\pi}|^2)$ or $O(x|\lambda_{K\pi}|)$.

One can observe that the $D^0 - \bar{D}^0$ mixing ($x$) and DCSD ($\lambda_{K\pi}$ or $\tilde{\lambda}_{K\pi}$) terms play insignificant roles in the decay rates $\text{R}(l^+, K^+\pi^-; T)_C$ and $\text{R}(l^-, K^-\pi^+; T)_C$. In contrast, $\text{R}(l^-, K^+\pi^-; T)_C$ and $\text{R}(l^+, K^-\pi^+; T)_C$ are remarkably suppressed due to the smallness of $x$ and $|\lambda_{K\pi}|$. Let us parametrize $\xi_{K\pi}$ and $\tilde{\xi}_{K\pi}$ as follows: $\xi_{K\pi} = |\xi_{K\pi}|e^{i(\delta_{K\pi}+\phi_t)}$ and $\tilde{\xi}_{K\pi} = |\xi_{K\pi}|e^{i(\delta_{K\pi}-\phi_t)}$, where $\delta_{K\pi}$ and $\phi_t$ stand for the strong phase shift and the weak transition phase respectively. Then $\text{Im}\lambda_{K\pi}$ and $\text{Im}\tilde{\lambda}_{K\pi}$ are given as

$$\begin{align}
\text{Im}\lambda_{K\pi} &= |\xi_{K\pi}|\sin(\delta_{K\pi} + \phi_t - 2\phi_m), \\
\text{Im}\tilde{\lambda}_{K\pi} &= |\xi_{K\pi}|\sin(\delta_{K\pi} - \phi_t + 2\phi_m).
\end{align} \tag{11}$$

In the standard model, we have $\phi_t = \arg[(V_{cd}V_{us}^*)/(V_{ud}V_{cs}^*)] \approx A^2\lambda^4\eta \leq 10^{-3}$ as well as $\phi_m = \arg[(V_{us}V_{cs}^*)/(V_{cs}V_{us}^*)]/2 \approx 0$, where $A$, $\lambda$ and $\eta$ are the Wolfenstein parameters. Since any new physics is unlikely to significantly affect the direct decays of $c$ quark, $\phi_t \approx 0$ is always a good approximation and will be used later on. New physics might introduce additional non-trivial phases into $\phi_m$ [10, 13], leading to observable $CP$-violating effects through the interplay of decay and mixing.

To isolate the mixing and DCSD effects, the following two types of measurables can be analyzed in experiments:

(a) The combined decay rate

$$\Omega_C^{+-}(T) \equiv \text{R}(l^-, K^+\pi^-; T)_C + \text{R}(l^+, K^-\pi^+; T)_C.$$

Explicitly, we get

$$\begin{aligned}\Omega_-^{+-}(T) &\propto 2\, e^{-T}\, [4|\xi_{K\pi}|^2 + x^2 T^2 + 4x|\xi_{K\pi}|T \sin\delta_{K\pi}\cos(2\phi_m)]\,, \\ \Omega_+^{+-}(T) &\propto 2\, e^{-T}\, [4|\xi_{K\pi}|^2 + x^2(2 + 2T + T^2) + 4x|\xi_{K\pi}|(1+T)\sin\delta_{K\pi}\cos(2\phi_m)]\,.\end{aligned} \quad (12)$$

Note that the strong phase shift $\delta_{K\pi}$ vanishes only in the limit of SU(3) symmetry [14]. To fit the CLEO II result for $D^0 \to K^\pm \pi^\mp$, which gives $|\xi_{K\pi}|^2 = (0.77 \pm 0.25 \pm 0.25)\%$ [15], one finds $\delta_{K\pi} \sim 5^0 - 13^0$ from a few phenomenological models [16]. For illustration, we show the time dependence of $\Omega_C^{+-}(T)$ in Fig. 1 by taking $|\xi_{K\pi}| = 0.08$, $\delta_{K\pi} = 10^0$, $\phi_m = 30^0$ and $x = 0.06$ (the upper bound of $|x|$ is 0.086 [12]). One can observe that $\Omega_-^{+-}(T)$ and $\Omega_+^{+-}(T)$ are most sensitive to the presence of DCSD or $D^0 - \bar{D}^0$ mixing in the regions $T \sim 1 - 4$ and $T \sim 0 - 4$, respectively. For small values of $T$, the signal of mixing in $\Omega_+^{+-}(T)$ is more significant than that in $\Omega_-^{+-}(T)$. Thus it is in principle favourable to study $D^0 - \bar{D}^0$ mixing by use of $(D^0 \bar{D}^0)_{C=+}$ events at the $\psi(4.14)$ resonance.

(b) The $CP$-violating asymmetry

$$\mathcal{A}_C^{+-}(T) \equiv \frac{\mathrm{R}(l^+, K^+\pi^-; T)_C - \mathrm{R}(l^-, K^-\pi^+; T)_C}{\mathrm{R}(l^+, K^+\pi^-; T)_C + \mathrm{R}(l^-, K^-\pi^+; T)_C}\,.$$

This signal arises from the interplay of DCSD and mixing. With the help of eqs. (9) and (10), we obtain

$$\begin{aligned}\mathcal{A}_-^{+-}(T) &\approx x|\xi_{K\pi}|T\cos\delta_{K\pi}\sin(2\phi_m)\,, \\ \mathcal{A}_+^{+-}(T) &\approx x|\xi_{K\pi}|(1+T)\cos\delta_{K\pi}\sin(2\phi_m)\,.\end{aligned} \quad (13)$$

It is clear that the above asymmetries are considerably suppressed due to the smallness of $x$ and $|\xi_{K\pi}|$. In addition, nonvanishing $\phi_m$ is a necessary condition to have nonzero $\mathcal{A}_C^{+-}(T)$. We illustrate the changes of $\mathcal{A}_C^{+-}(T)$ with $T$ in Fig. 2, where the inputs are the same as in case (a). Here it should be noted that large data samples for $(l^+, K^+\pi^-)$ and $(l^-, K^-\pi^+)$ events are available at a $\tau$-charm factory, since such joint decay rates are not suppressed by small mixing and DCSD effects (see eqs. (9) and (10)). Hence it is quite likely to measure $\mathcal{A}_C^{+-}(T)$ in the near future, if the magnitudes of $x$ and $\phi_m$ are large enough (e.g., $x \geq 0.05$ and $\sin(2\phi_m) \geq 0.5$).

A natural question to be asked is if the decay-time differences $T$ (or $t_-$) between the semileptonic and nonleptonic $D$ decays can really be measured on the $\psi(3.77)$ and $\psi(4.14)$ resonances. For symmetric $e^+e^-$ collisions at the $\Upsilon(4S)$ resonance, the produced $B_d^0 \bar{B}_d^0$ pair is almost at rest so that their mean decay length is insufficient for identifying the decay vertices or measuring the decay-time difference [11]. Hence designing an asymmetric $B$ factory is necessary to study the time distribution of joint $(B_d^0 \bar{B}_d^0)$ decays and $CP$ violation. For the similar kinetic reasons, a symmetric $\tau$-charm factory might also be unable to resolve the decay vertices of $(D^0\bar{D}^0)$ events, and this would make the time-dependent measurements discussed above impossible in practice. Of course, an instructive analysis of this problem is desirable to give definite conclusions for the $\tau$-charm factory designers. From our point of view it is suggestive to have an asymmetric $\tau$-charm factory running at the $\psi(3.77)$ and $\psi(4.14)$ resonances, which can allow various possible measurements of $D^0 - \bar{D}$ mixing, DCSD and $CP$-violating effects in coherent $(D^0\bar{D}^0)_C$ decays.

## 4. Time integration of the decay rates

In this section we calculate the time-integrated probabilities of the joint decays $(D^0\bar{D}^0)_C \to (l^\pm X^\mp)(K^\pm\pi^\mp)$. Without loss of any generality, we start from eqs. (4) and (5), where no special approximation has been made. Integrating R($l^\pm, t_1; K^\pm\pi^\mp, t_2)_C$ over $t_1 \in [0, +\infty)$ and $t_2 \in [0, +\infty)$, one obtains:

$$\begin{aligned}
\mathrm{R}(l^+, K^+\pi^-)_C &\propto |A_l|^2 |A_{K\pi}|^2 \left[ \frac{1+Cy^2}{(1-y^2)^2}(1+|\lambda_{K\pi}|^2) - \frac{2(1+C)y}{(1-y^2)^2}\mathrm{Re}\lambda_{K\pi} \right. \\
&\left. + \frac{1-Cx^2}{(1+x^2)^2}(1-|\lambda_{K\pi}|^2) - \frac{2(1+C)x}{(1+x^2)^2}\mathrm{Im}\lambda_{K\pi} \right] ,
\end{aligned} \quad (14a)$$

$$\begin{aligned}
\mathrm{R}(l^-, K^-\pi^+)_C &\propto |A_l|^2 |A_{K\pi}|^2 \left[ \frac{1+Cy^2}{(1-y^2)^2}(1+|\tilde{\lambda}_{K\pi}|^2) - \frac{2(1+C)y}{(1-y^2)^2}\mathrm{Re}\tilde{\lambda}_{K\pi} \right. \\
&\left. + \frac{1-Cx^2}{(1+x^2)^2}(1-|\tilde{\lambda}_{K\pi}|^2) - \frac{2(1+C)x}{(1+x^2)^2}\mathrm{Im}\tilde{\lambda}_{K\pi} \right] ;
\end{aligned} \quad (14b)$$

and

$$\begin{aligned}
\mathrm{R}(l^-, K^+\pi^-)_C &\propto |A_l|^2 |A_{K\pi}|^2 e^{-4\mathrm{Im}\phi_m} \left[ \frac{1+Cy^2}{(1-y^2)^2}(1+|\lambda_{K\pi}|^2) - \frac{2(1+C)y}{(1-y^2)^2}\mathrm{Re}\lambda_{K\pi} \right. \\
&\left. - \frac{1-Cx^2}{(1+x^2)^2}(1-|\lambda_{K\pi}|^2) + \frac{2(1+C)x}{(1+x^2)^2}\mathrm{Im}\lambda_{K\pi} \right] ,
\end{aligned} \quad (15a)$$

$$\begin{aligned}
\mathrm{R}(l^+, K^-\pi^+)_C &\propto |A_l|^2 |A_{K\pi}|^2 e^{+4\mathrm{Im}\phi_m} \left[ \frac{1+Cy^2}{(1-y^2)^2}(1+|\tilde{\lambda}_{K\pi}|^2) - \frac{2(1+C)y}{(1-y^2)^2}\mathrm{Re}\tilde{\lambda}_{K\pi} \right. \\
&\left. - \frac{1-Cx^2}{(1+x^2)^2}(1-|\tilde{\lambda}_{K\pi}|^2) + \frac{2(1+C)x}{(1+x^2)^2}\mathrm{Im}\tilde{\lambda}_{K\pi} \right] ,
\end{aligned} \quad (15b)$$

where $y \equiv \Delta\Gamma/(2\Gamma)$ is another $D^0 - \bar{D}^0$ mixing parameter. ¿From the above formulas one can observe that for $C = -$ case the interference terms $\mathrm{Re}\lambda_{K\pi}$, $\mathrm{Im}\lambda_{K\pi}$, $\mathrm{Re}\tilde{\lambda}_{K\pi}$ and $\mathrm{Im}\tilde{\lambda}_{K\pi}$ disappear in the joint decay rates. Indeed this is a generic feature of coherent $(D^0\bar{D}^0)_{C=-}$ decays, independent of the final products to be semileptonic or nonleptonic states [11].

In the assumption of $|\Delta\Gamma| << |\Delta m|$, one equivalently has $|y| << |x|$. Subsequently we neglect the contributions from $\mathrm{Im}\phi$ and $y$ as well as the $O(\leq x^3)$ terms, in order to simplify eqs. (14) and (15). The relevant time-integrated decay rates turn out to be

$$
\begin{aligned}
\text{R}(l^+, K^+\pi^-)_- &= \text{R}(l^-, K^-\pi^+)_- \propto |A_l|^2 |A_{K\pi}|^2 \left[ 2 - x^2 \left(1 - |\lambda_{K\pi}|^2\right) \right] , \\
\text{R}(l^-, K^+\pi^-)_- &= \text{R}(l^+, K^-\pi^+)_- \propto |A_l|^2 |A_{K\pi}|^2 \left[ 2|\lambda_{K\pi}|^2 + x^2 \left(1 - |\lambda_{K\pi}|^2\right) \right] ;
\end{aligned}
\qquad (16)
$$

and

$$
\begin{aligned}
\text{R}(l^+, K^+\pi^-)_+ &\propto |A_l|^2 |A_{K\pi}|^2 \left[ 2 - 3x^2 \left(1 - |\lambda_{K\pi}|^2\right) - 4x\,\text{Im}\lambda_{K\pi} \right] , \\
\text{R}(l^-, K^-\pi^+)_+ &\propto |A_l|^2 |A_{K\pi}|^2 \left[ 2 - 3x^2 \left(1 - |\tilde{\lambda}_{K\pi}|^2\right) - 4x\,\text{Im}\tilde{\lambda}_{K\pi} \right] , \\
\text{R}(l^-, K^+\pi^-)_+ &\propto |A_l|^2 |A_{K\pi}|^2 \left[ 2|\lambda_{K\pi}|^2 + 3x^2 \left(1 - |\lambda_{K\pi}|^2\right) + 4x\,\text{Im}\lambda_{K\pi} \right] , \\
\text{R}(l^+, K^-\pi^+)_+ &\propto |A_l|^2 |A_{K\pi}|^2 \left[ 2|\tilde{\lambda}_{K\pi}|^2 + 3x^2 \left(1 - |\tilde{\lambda}_{K\pi}|^2\right) + 4x\,\text{Im}\tilde{\lambda}_{K\pi} \right] .
\end{aligned}
\qquad (17)
$$

Here the $x^2|\lambda_{K\pi}|^2$ and $x^2|\tilde{\lambda}_{K\pi}|^2$ terms are further negligible. These formulas can be compared with those given in eqs. (9) and (10).

In practical experiments, the combined decay rates

$$
\Omega_C^{+-} \equiv \text{R}(l^-, K^+\pi^-)_C + \text{R}(l^+, K^-\pi^+)_C
$$

should be observable. Similar to eq. (12), we find

$$
\begin{aligned}
\Omega_-^{+-} &\propto 2|A_l|^2 |A_{K\pi}|^2 \left[ 2|\xi_{K\pi}|^2 + x^2 \left(1 - |\xi_{K\pi}|^2\right) \right] \\
\Omega_+^{+-} &\propto 2|A_l|^2 |A_{K\pi}|^2 \left[ 2|\xi_{K\pi}|^2 + 3x^2 \left(1 - |\xi_{K\pi}|^2\right) + 4x|\xi_{K\pi}| \sin\delta_{K\pi} \cos(2\phi_m) \right] .
\end{aligned}
\qquad (18)
$$

Clearly the effects of $D^0$–$\bar{D}^0$ mixing and DCSD cannot be distinguished from each other, if the sizes of $|x|$ and $|\xi_{K\pi}|$ are comparable. In the case of $|x| \ll |\xi_{K\pi}|$, the relation $\Omega_-^{+-} = \Omega_+^{+-} \propto |\xi_{K\pi}|^2$ holds, as expected in the standard model. Here the time-integrated $CP$ asymmetries can be defined as

$$
\mathcal{A}_C^{+-} \equiv \frac{\text{R}(l^+, K^+\pi^-)_C - \text{R}(l^-, K^-\pi^+)_C}{\text{R}(l^+, K^+\pi^-)_C + \text{R}(l^-, K^-\pi^+)_C} .
$$

Explicitly, we obtain

$$
\begin{aligned}
\mathcal{A}_-^{+-} &\approx 0 , \\
\mathcal{A}_+^{+-} &\approx 2x|\xi_{K\pi}| \cos\delta_{K\pi} \sin(2\phi_m) .
\end{aligned}
\qquad (19)
$$

This result is quite similar to that obtained for $B_d^0$ vs $\bar{B}_d^0 \to D^{\pm}\pi^{\mp}$ at the $\Upsilon(4S)$ [11].

Now we roughly estimate the magnitudes of $\Omega_-^{+-}$, $\Omega_+^{+-}$ and $\mathcal{A}_+^{+-}$, to give one a feeling of ballpark numbers to be expected. Assuming the semileptonic decay mode serving for flavor tagging to be $D^0 \to K^- e^+ \nu_e$, we have its branching ratio $\text{Br}(D^0 \to K^- e^+ \nu_e) \approx 3.8\%$ [12]. In addition, the current data give $\text{Br}(D^0 \to K^-\pi^+) \approx 4.01\%$ [12]. Then $\text{R}(l^+, K^+\pi^-)_\pm$ and $\text{R}(l^-, K^-\pi^+)_\pm$ are at the level $10^{-3}$ or so, while $\Omega_\pm^{+-}$ may be of the order $10^{-5}$ if we input $x \sim 0.06$. Within the experimental capabilities of a $\tau$-charm factory, it is possible to measure $\Omega_\pm^{+-}$ as well as $\Omega_\pm^{+-}(T)$ to an acceptable degree of accuracy with about $10^7$ ($D^0\bar{D}^0$) events. Furthermore, the upper bound of the $CP$ asymmetry $\mathcal{A}_+^{+-}$ can be obtained by use of the experimental results $|x| < 0.086$ and $|\xi_{K\pi}| \approx 0.088$ [12, 15]. Taking $\cos\delta_{K\pi} = 1$ and $\sin(2\phi_m) = \pm 1$, we get $|\mathcal{A}_+^{+-}| < 0.015$. Similar constraints on the time-dependent $CP$ asymmetries in eq. (13) are then obtainable through the relations $\mathcal{A}_-^{+-}(T) = 0.5 A_+^{+-} T$ and $\mathcal{A}_+^{+-}(T) = 0.5 A_+^{+-}(1+T)$. In the assumption of perfect detectors or 100% tagging efficiencies, one needs about $10^8$ $(D^0 \bar{D}^0)$ events to uncover $|\mathcal{A}_+^{+-}| \sim 0.01$ at the level of three standard deviations or to measure $|\mathcal{A}_+^{+-}| \sim 0.005$ at the level of one standard deviation. Accumulation of so many events is of course a serious challenge to all types of facilities for charm physics, but it should be achieved in the second-round experiments of a $\tau$-charm factory.

## 5. Further discussions

Taking into account the $D^0 - \bar{D}^0$ mixing and DCSD effects, we have calculated the time-dependent and time-integrated decay rates for neutral $D$ decays to $K^\pm \pi^\mp$ at a $\tau$-charm factory. The similarities and differences between the decay modes from $(D^0 \bar{D}^0)_{C=+}$ and $(D^0 \bar{D}^0)_{C=-}$ events are illustrated. We have also taken a look at the $CP$ asymmetry of $D^0$ vs $\bar{D}^0 \to K^\pm \pi^\mp$, which is induced by the interplay of $D^0 - \bar{D}^0$ mixing and DCSD. A similar discussion can be given for other neutral $D$ decays to two-body non-$CP$ eigenstates with DCSD effects included, such as $D^0 \to K^\pm \rho^\mp, K^{*\pm} \pi^\mp$ and their $CP$-conjugate processes. These channels may occur through the same weak interactions, but their final-state strong interactions should be different from one another (e.g., $\delta_{K\pi} \neq \delta_{K\rho}$). If the SU(3) breaking effects in $D^0 \to (K^\pm, K^{*\pm}) + (\pi^\mp, \rho^\mp, a_1^\mp, \text{etc})$ are not so significant that all the strong phase shifts lie in the same quadrant as $\delta_{K\pi}$, then a sum over these modes is possible to increase the number of decay events in statistics, with few dilution effect on the signal of $CP$ violation.

It has been pointed out that the coherent decays $(D^0 \bar{D}^0)_C \to (K^\pm \pi^\mp)(K^\pm \pi^\mp)$ can be used to search for $D^0 - \bar{D}^0$ mixing and to separate it from the DCSD effect [1, 17]. The relevant measurables may be

$$r_C^{+-} \equiv \frac{\text{R}(K^+\pi^-, K^+\pi^-)_C}{\text{R}(K^-\pi^+, K^+\pi^-)_C} \quad \text{or} \quad r_C^{-+} \equiv \frac{\text{R}(K^-\pi^+, K^-\pi^+)_C}{\text{R}(K^-\pi^+, K^+\pi^-)_C} .$$

Since in previous calculations the effects of $CP$ violation and nonvanishing $\delta_{K\pi}$ on $r_C^{\pm\mp}$ were neglected, it is worth having a recalculation of these observables without special approximations. Explicitly, we find [3]:

$$r_-^{+-} \approx r_-^{-+} \approx \frac{1}{2}\left(x^2 + y^2\right) , \tag{20}$$

and

$$\begin{aligned} r_+^{+-} &\approx \tfrac{3}{2}(x^2+y^2) + 4(|\lambda_{K\pi}|^2 + x\text{Im}\lambda_{K\pi} - y\text{Re}\lambda_{K\pi}) , \\ r_+^{-+} &\approx \tfrac{3}{2}(x^2+y^2) + 4(|\tilde{\lambda}_{K\pi}|^2 + x\text{Im}\tilde{\lambda}_{K\pi} - y\text{Re}\tilde{\lambda}_{K\pi}) . \end{aligned} \tag{21}$$

---
[3] Here it is not necessary to assume $|x| \gg |y|$, but $\text{Im}\phi_m = 0$ has been used.

One can see that $r_+^{+-} \neq r_+^{-+}$ in general, and their difference implies the presence of $CP$ violation:

$$r_+^{-+} - r_+^{+-} \approx 8|\xi_{K\pi}|\sin(2\phi_m)(x\cos\delta_{K\pi} + y\sin\delta_{K\pi}) \ . \tag{22}$$

In the approximation of $y = 0$, we get $r_+^{-+} - r_+^{+-} \approx 4\mathcal{A}_+^{+-}$. Of course, this $CP$ asymmetry is very interesting and should be searched for in experiments.

Certainly there are other possibilities to explore $CP$ violation in neutral $D$ decays [1, 2]. In particular, direct $CP$ violation in the decay amplitude is expected to give rise to observable effects in some decay modes to $CP$ eigenstates (e.g., $D^0$ vs $\bar{D} \to K^+K^-, \pi^+\pi^-$ and $K_S\rho$) [3, 18]. Following the same approaches outlined in this work, a detailed analysis of the time-dependent and time-integrated decay rates for $(D^0\bar{D}^0)_C \to (K^+K^-)(l^\pm X^\mp)$ etc is worth while. In practical experiments, however, a symmetric $\tau$-charm factory seems difficult to implement the time-dependent measurements. Hence one is invited to speculate an asymmetric machine and its physics potential [19].

Our conclusion is that coherent $(D^0\bar{D}^0)$ decays to $(K^\pm\pi^\mp)(l^\pm X^\mp)$ and $(K^\pm\pi^\mp)(K^\pm\pi^\mp)$ contain clear signals of $D^0 - \bar{D}^0$ mixing, DCSD and $CP$ violation. They can be measured through either time-dependent or time-integrated way (or both) at a $\tau$-charm factory running at the $\psi(3.77)$ or $\psi(4.14)$ resonance. It is also interesting to study a variety of decay modes similar to $D^0 \to K^\pm\pi^\mp$ at the same experimental scenario.

I would like to thank Harald Fritzsch for his warm hospitality and constant encouragement. This work was stimulated from useful discussions with Dongsheng Du, Daniel M. Kaplan and Tiehui (Ted) Liu. I am greatly indebted to Ted for his reading the manuscript and giving many constructive comments on it. Finally I acknowledge the Alexander von Humboldt Foundation of Germany for its financial support.

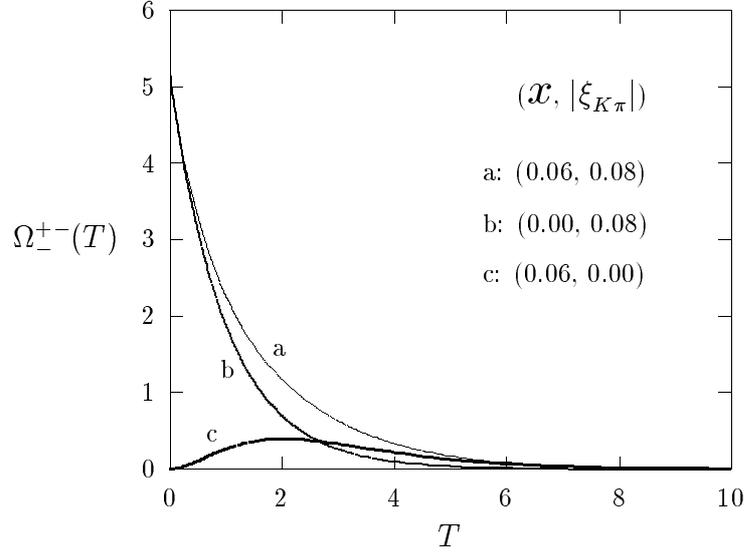

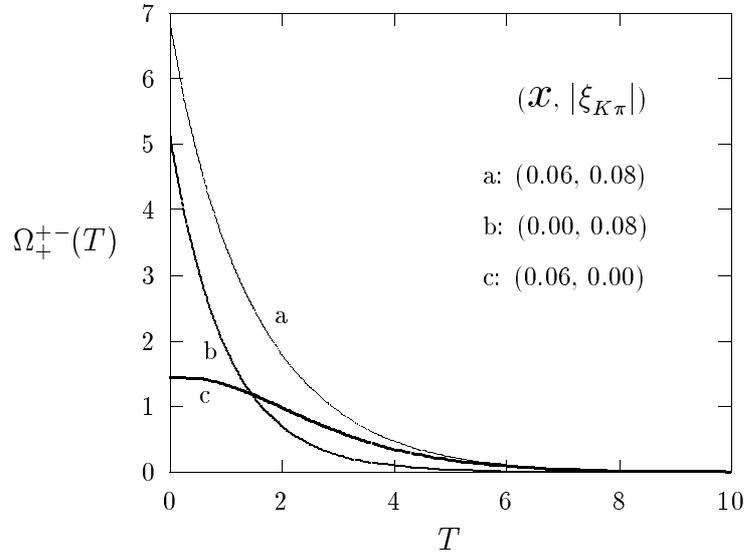

Figure 1: Illustrative plots of the time dependence of $\Omega_-^{+-}(T)$ and $\Omega_+^{+-}(T)$ (in common but arbitrary units), where $\delta_{K\pi} = 10^0$ and $\phi_m = 30^0$ are taken.

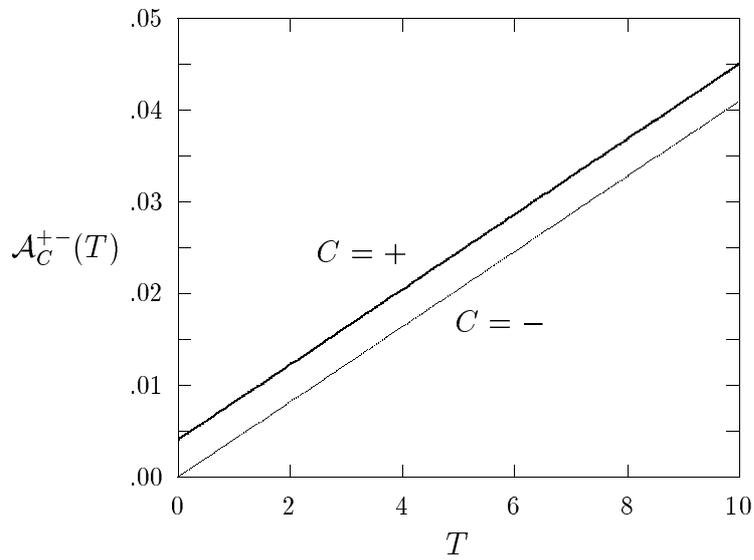

Figure 2: Illustrative plot of the time dependence of $\mathcal{A}_C^{+-}(T)$, where $x = 0.06$, $|\xi_{K\pi}| = 0.08$, $\delta_{K\pi} = 10^0$ and $\phi_m = 30^0$ are taken.